\documentclass[prb,twocolumn,aps,amsmath,amssymb,showpacs,superscriptaddress]{revtex4}

\usepackage{graphicx}
\usepackage{dcolumn}
\usepackage{bm}
\usepackage[T1]{fontenc}

\begin{document}

\title{Current-Phase Relation and Josephson Inductance of Superconducting Cooper Pair Transistor}
\author{A.~Paila}\email[Corresponding author. \\ Electronic address:
pjh@boojum.hut.fi]{}
\author{D.~Gunnarsson}
\affiliation{Low Temperature Laboratory, Helsinki University of Technology, Espoo, Finland}
\author{J.~Sarkar}
\affiliation{Low Temperature Laboratory, Helsinki University of Technology, Espoo, Finland}
\author{M.~A.~Sillanp\"a\"a}
\affiliation{Low Temperature Laboratory, Helsinki University of Technology, Espoo, Finland}
\author{P.~J.~Hakonen}
\affiliation{Low Temperature Laboratory, Helsinki University of Technology, Espoo, Finland}

\begin{abstract}

We have investigated the Josephson inductance $L_J$ of
a Superconducting Cooper Pair Transistor (SCPT). We traced the
inductance using microwave reflection measurements on a
tuned resonance circuit in which a SCPT was mounted in parallel to
a $ \sim 200$ pH strip line inductance. When the inverse of the
Josephson inductance,  determined on the charge-phase bias plane
for a SCPT with a Josephson to Coulomb energy ratio of $E_J/E_C
=1.75$, is integrated over the phase, we obtain a current-phase
relation, which is strongly non-sinusoidal near the charge
degeneracy point.

\end{abstract}

\pacs{85.25.Cp, 85.35.Gv, 74.78.Na, 74.25.Fy}

\maketitle Josephson effect is one of the most spectacular
phenomena in quantum coherent matter.  In addition to
superconductors, it has been studied in superfluids
\cite{AvenelPRL88,DavisRMP} and Bose-Einstein condensates
\cite{Kasevich98,Cataliotti2001}. The coupling between the ordered
states across a Josephson tunnel junction can be described by the<
Josephson energy $E(\varphi)$ where $\varphi$ denotes the phase
difference between the order parameter fields on the opposite
sides of the junction. Equivalently, the nature of a Josephson
junction can be characterized by its current-phase relation (CPR)
$I(\varphi)=\frac{2\pi}{\Phi_0}\frac{d E}{d\varphi}$
where $\Phi_0=\frac{h}{2e}$ is the flux quantum. In a
regular tunnel barrier junction, the CPR is sinusoidal, which
leads to a non-linear dynamical inductance of
$L_J^{-1}=\frac{d I}{d \Phi} = \frac{2\pi}{\Phi_0}
I_c \cos \varphi$ where $\Phi=\Phi_0 \varphi/2\pi$ is the flux
corresponding to the phase $\varphi$ \cite{LikharevBOOK}. In
junctions with sinusoidal CPR, $|L_J|$ is $\pi$-periodic and the
inductance changes sign at $\varphi=\pi/2$ and at
$\varphi=3\pi/2$. As the inverse inductance describes the energy
exchange rate of the junction with its surrounding electrical
circuit, we observe that the junction varies periodically between
a circuit element either receiving or releasing energy, in a
sinusoidal fashion in this case. In different kinds of junctions,
for example, in superconducting quantum point contacts (QPC) and
in diffusive SNS junctions, the CPR can be strongly non-sinusoidal
\cite{GolubovRMP}. In the outmost case, like in a QPC with near
perfect transmission ($\tau \rightarrow 1$), $I(\varphi)$ becomes
discontinuous \cite{Kulik77,BeenakkerHouten} and the region of
negative Josephson inductance shrinks to zero. Similar discontinuous behavior takes place also in two symmetric, classical junctions in series \cite{MatveevPRL93}.

In a mesoscopic double junction device, charge
quantization on the island between the junctions starts to play a
role. This leads to phase fluctuations across the individual
junctions, even if the total phase bias acts like a classical
variable and can be fixed. As a consequence, the CPR is strongly modified from the classical result. The energy bands, calculable from the Schr\"odinger equation with
the Coulombic term acting as the kinetic energy operator in the
phase picture \cite{LikharevBOOK,AverinLikharev}, are well known and
these eigenenergies form the cornerstone of charge qubits as well
as charge-phase qubits \cite{Nakamura99,Vion02,MSS}. A lot of
energy spectroscopy has been performed on these energy levels
which, however, does not probe the current-phase relation of the
ground state unless assumptions on the symmetry properties of the
bands are made. Here we report measurements on the Josephson
inductance of a superconducting Cooper pair
transistor and determine its CPR. We present, for the first
time to our knowledge, results on the current-phase
relation that demonstrate the
strongly non-sinusoidal character near the charge degeneracy point. Our results are based
on accurate reflection measurements of a charge-phase qubit
coupled to a microwave resonator.

In the superconducting state, a tunnel junction stores energy in the form of Josephson coupling energy according to $E=-E_J \cos{(\varphi)}$, where the Josephson energy $E_J$ is related to the junction critical current $I_C$ through $I_C = 2eE_J/\hbar$. According to electromagnetism, the inductance is defined by the differential of the total energy
\begin{equation}\label{L}
 L=\left( \frac{\mathrm{\partial}^2 \mathcal{H}}{\mathrm{\partial}\Phi^2} \right)^{-1}=\left(\frac{\Phi_0}{2\pi}\right)^2 \left( \frac{\mathrm{\partial}^2 \mathcal{H}}{\mathrm{\partial}\varphi^2}, \right)^{-1}.
\end{equation}
In the regime of small fluctuations, $\varphi \ll 1$, Eq. (\ref{L}) yields for a single classical Josephson junction (JJ) a parametric inductance $L_J^0 = \hbar/(2eI_C^0)$. For arbitrary phase bias, one gets $L_J^{-1} = [L_J^0]^{-1}\cos\varphi$.

Although typically applied to single JJs, these concepts can be
applied to general junction combinations like a single Cooper-pair
transistor (SCPT) which consists of two JJs and one capacitively
coupled gate electrode. The full description of SCPT's energy
storing capabilities is governed by its Hamiltonian
\cite{AverinLikharev}
\begin{equation}\label{ham}
 \hat{H} = E_C (\hat{n}-n_g)^2 -2E_J \cos{\frac{\varphi}{2}}\cos{\hat{\theta}}+2dE_J\sin{\frac{\varphi}{2}}\sin{\hat{\theta}},
\end{equation}
where $E_C=e^2/(2C)$ is the charging energy of the island, $\hat
n$ denotes the number of extra electron charges on the island, and
$n_{\rm g} = C_{\rm g} V_{\rm g} / e$ is gate-voltage-induced
charge in units of single electron on the gate with capacitance
$C_{\rm g}$.
 $\hat n$ is conjugate to $\hat\theta/2$, where $\hat\theta$ is
the superconducting phase on the island. The asymmetry of the two Josephson
 junctions of the SCPT is described by
 $d=(E_{{\rm J}1}-E_{{\rm J}2})/2E_{\rm J}$.
As we are interested in the phase response, the electrostatic
energy associated with the gate capacitance may be neglected here,
contrary to the case in Ref. \onlinecite{MikaQcap} where
Josephson capacitance was considered. In order to get the
Josephson inductance, the eigenenergies of Hamiltonian of Eq.
(\ref{ham}) have to be solved first \cite{Zorin1997,Zorin2002}.

Qualitative analysis can be performed in the charge basis taking
only two charge states into account. Then, the Hamiltonian can be
written in terms of Pauli spin matrices as \cite{MSS}
\begin{equation}
 \hat{H} = -\frac{1}{2}\left(B_x \hat{\sigma}_x + B_y \hat{\sigma}_y + B_z \hat{\sigma}_z \right),
\end{equation}
where the magnetic field components are $B_z=4E_C(1-n_g)$, $B_x=2E_J\cos{(\varphi/2)}$ and $B_y=2dE_J\cos{(\varphi/2)}$.
By differentiating the energy eigenvalues, one obtains an analytic form for the inductance of the ground state
\begin{equation}\label{L2level}
\frac{L_{J\downarrow}}{(\Phi_0/2\pi)^2}=
\frac{\frac{8\sqrt{2}}{E_J}\left(1+
\frac{8}{\gamma^2}\left(1-n_g\right)^2+\cos\varphi\right)^{3/2}}
{\left(4+\frac{32}{
\gamma^2}\left(1-n_g\right)^2\right)\cos\varphi+3+\cos
2\varphi},
\end{equation}
where $\gamma$ denotes the ratio $\gamma=E_J/E_c$. Because of the symmetry in the two level approximation one gets
$L_{J\uparrow}=-L_{J\downarrow}$ for the excited state. When $\gamma
\ll 1$, the two level approximation describes the energy bands
accurately but, when $\gamma \sim 1$, more charge states are
needed. $L_{J\downarrow}$ calculated from the two level
approximation, compared to the measured data in Fig. \ref{fig:fit}
below, is found to fall short and numerical calculations including more levels are
needed to make a detailed comparison. Nevertheless, Eq.
(\ref{L2level}) reveals clearly the main features: 1) the
supercurrent is strongly suppressed with lowering $\gamma$, and 2) the
non-harmonic character grows when approaching the charge
degeneracy point \cite{Zorin2002}.

\begin{figure}
 \centering
 \includegraphics[width=0.8\columnwidth]{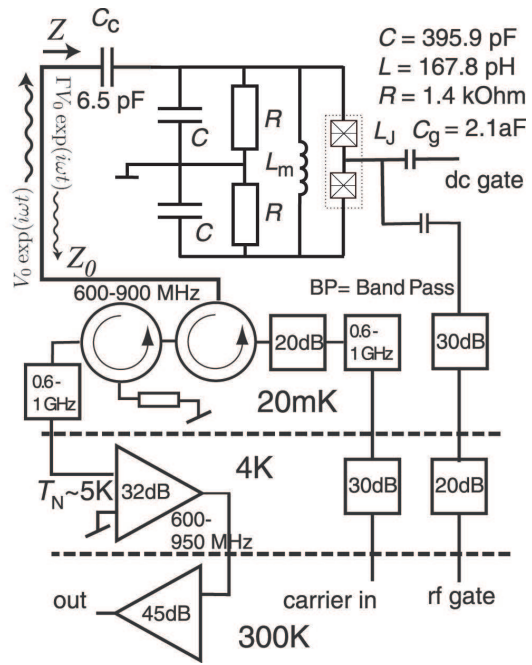}
 \caption{The circuit diagram of our symmetrized L-SET device. Here the SCPT's Josephson inductance $L_J$ is in parallel with an on-chip microstrip inductor $L_m$ which form together the inductor in the LC-oscillator: $1/L=1/L_J+1/L_m$. $C_c$ is the coupling capacitor that decouples the resonator from the driving signal generator.}
 \label{fig:setup}
\end{figure}

In the experiment, the SCPT's reactive response was measured in the
L-SET (inductively read single-electron transistor) configuration
\cite{LSET}, where the junctions are embedded in an electric
LC-oscillator circuit (see also Refs.
\onlinecite{IlichevRSI,Aumentado06,ShevchenkoEPJB}). The L-SET sample was
fabricated using a standard e-beam lithography process followed by
evaporation of  aluminum films at two different angles of
incidence. Circuit diagram of the L-SET sample is shown in Fig.
\ref{fig:setup}. The SCPT is attached in parallel with an inductor
that is realized using an on-chip microstrip loop. The length of
the loop is 300 $\mu$m, which yields $L_m = 168$ pH inductance.
The two parallel capacitors of the resonator are made by
patterning two large plates ($1.3 \mathrm{mm} \times 1.3
\mathrm{mm}$) on the substrate (see Fig. 2 of Ref.
\onlinecite{Gunnarsson}). The bottom plate of the capacitor is
formed by a 150 nm thick niobium ground plane layer under the
$Si_3 N_4$ dielectric layer whose thickness is around 300 nm. The
measured resonance frequency yields for the capacitors  $C = 396$
pF, which agrees well with the parallel plate result using
$\epsilon_r = 7.5$ for $Si_3 N_4$. Because our capacitors are
non-ideal, we have to take into account their ac-losses by
introducing a shunting resistance whose value $R=1.4
~\mathrm{k\Omega}$ was determined from the \emph{Q}-factor of the
resonator. The coupling capacitor $C_c=6.5$ pF, that decouples the
oscillator from the driving coaxial cable, is a separate surface
mount capacitor that is bonded with aluminum wires to the top pad
of one of the capacitors. The gate modulation was 2e periodic and
from the period we deduce $C_g= 2.1$ aF. From the rf-spectroscopy
we conclude that $E_c/k_B = 0.8 ~\mathrm{K}$, $E_J/k_B \approx 1.6
~\mathrm{K}$, and the asymmetry parameter $d \approx 0.06$.

\begin{figure}
 \centering
 \includegraphics[width=0.8\columnwidth]{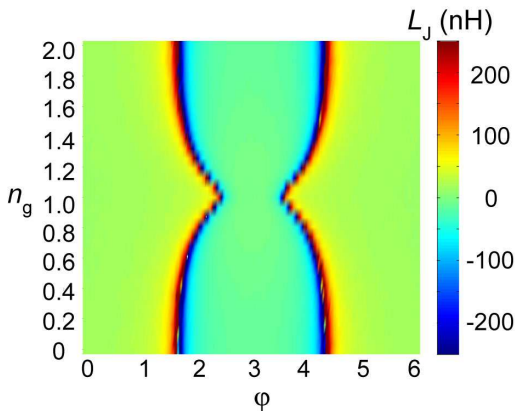}
 \caption{Measured inductance of the ground state on the phase bias vs. gate charge plane (inductance scale limited to $\pm 250 ~\mathrm{nH}$). The plot was obtained using Eq. \ref{eq:J_ind_exp} and the data scans presented in the supplementary material.}
 \label{fig:ground_ind_exp}
\end{figure}

The Josephson inductance was measured using a microwave reflection measurement setup working around $f_m=870$ MHz. Continuous microwave signal was sent from one port of a vector network analyzer through a series of attenuators and a circulator to the sample. Part of the sent signal was reflected back to the transmission line due to the impedance mismatch between $Z_0=50 ~\Omega$ and the impedance of the resonator circuit, $Z$ (see Fig. \ref{fig:setup}): $V_{out}=\Gamma V_{in}$, where $\Gamma = (Z-Z_0)/(Z+Z_0)$. The reflected signal was then taken from the third port of the circulator (isolation 18 dB), amplified with an array of cold and room temperature amplifiers and connected to the second channel of the network analyzer. The network analyzer gave us the magnitude signal and the phase argument (See Figs. 6 and 7 in Appendix I.) of the reflection coefficient which were used to calculate the effective impedance of the oscillator circuit. Using standard circuit analysis with linear elements, the Josephson inductance then can be calculated from the expression
\begin{equation}\label{eq:J_ind_exp}
 L_J=\frac{1}{i\omega}\left[ \left( \left( \frac{1}{ Z-Z_C} - \frac{1}{Z_R} \right)^{-1}-Z_R \right)^{-1}-\frac{1}{Z_L}\right]^{-1},
\end{equation}
where shorthand notations have been assigned to impedances of the
circuit components $Z_C = (i\omega C_c)^{-1} $, $Z_L = i\omega L$,
and $Z_R = (1/R+i\omega C)^{-1}$. The Josephson inductance,
obtained by using the experimentally determined $Z$, is shown in
Fig. \ref{fig:ground_ind_exp}. Ideally, the zero phase of the
reflection signal should be determined separately for each bias
point by plotting $\Gamma(f)$ parametrically on the Smith chart
and looking for the symmetry axis of the resulting circle (see,
e.g. Ref. \onlinecite{Kajfez}). We, however, did this check only at one
bias point and relied that the reference phase will not change
substantially. The reflection magnitude scale was fixed using
reflected signal far away from the resonance and the known circuit
elements. The finite measurement signal averages the energy bands over phase direction. The power used in measurement, -129 dBm, corresponds to phase amplitude of 0.24 rad (p-p) and causes a 3 percent error to the inductance.

\begin{figure}
 \centering
 \includegraphics[width=0.8\columnwidth]{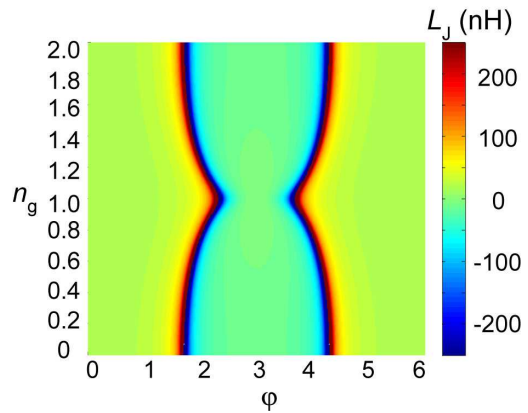}
 \caption{Theoretically calculated $L_J$ of the ground state on the $\varphi$ vs. $n_g$ plane ($|L_J|$ < $\pm 250 ~\mathrm{nH}$). The eigenvalues of the SCPT were numerically calculated using 5 charge states.}
 \label{fig:ground_ind_theory}
\end{figure}

The corresponding inductance plot obtained from theory is shown in
Fig. \ref{fig:ground_ind_theory}; five discrete charge states were
used in the determination of the eigenvalues of Eq. (\ref{ham}).
The general agreement between Figs. \ref{fig:ground_ind_exp} and
\ref{fig:ground_ind_theory} is remarkable, except near the charge
degeneracy point. At $n_g =1$, the gap between the ground and
excited states becomes small, about $\Delta/h = 3.8$ GHz, and the
upper state can  either be thermally excited at our effective
noise temperature of $\sim 40$ mK, governed by the thermalization
of our circulators, or by multiphoton processes due to the
microwave drive at 872 MHz. In this case, the effective inductance
is a mixture of the curvatures of the two energy levels making the
analysis  difficult. The mixing can, however,  be used to
determine the energy level occupation in systems where SCPT is
considered as a qubit\cite{LZprl}.
%
%

\begin{figure}
 \centering
 \includegraphics[width=0.9\columnwidth]{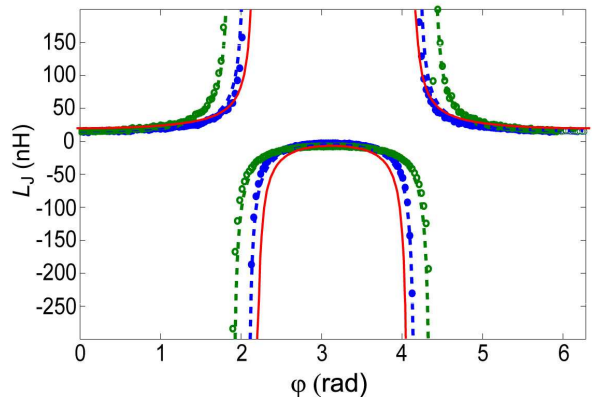}
 \caption{Comparison of the measured Josephson inductance and the theoretical curves at two charge bias values 1.25e (filled/blue) and 1.51e (open/green)).  The red solid line illustrates $L_J$ obtained from the two level approximation.}
 \label{fig:fit}
\end{figure}

A more quantitative comparison between the experimental data and
the theoretical prediction for $L_J(\varphi)$ is shown in Fig.
\ref{fig:fit}. The theory models well the experimental result when
$E_J/k_B = 1.5$ K is employed. The SCPT behaves as a non-linear
inductor whose inductance is tunable by phase bias from a few nH
to infinity (in ideal case). There exist substantial regions in
the charge-phase plane where the Josephson inductance takes
negative values, and the negative-value regime is diminished as
the charge degeneracy point is approached.

\begin{figure}[t]
 \centering
 \includegraphics[width=7.5cm]{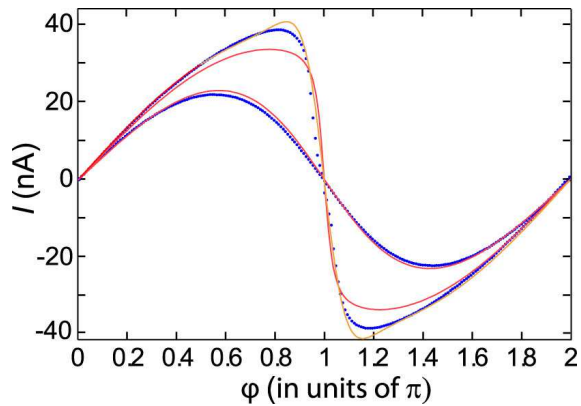}
 \caption{Current-phase relations (CPR) obtained by integration of the inverse Josephson inductance at two gate charge values: 0 (nearly sinusoidal) and $1 e$ (strongly non-sinusoidal).  Experimental curves are denoted by blue dots while the red solid curves display the CPRs calculated from Eq. (\ref{ham}); the orange solid curve depicts the result calculated from Eq. (\ref{TOTcurr}) using $T=0.28$ K.}
 \label{fig:CPR}
\end{figure}

The current-phase relation can be obtained by integrating the
inverse of Josephson inductance: The relation
$L_J^{-1}=\frac{\partial I}{\partial \Phi}$ yields $ I(\varphi) =
\frac{\Phi_0}{2\pi}\int_0^\varphi L_J^{-1} d\varphi + g(V_g)$
where the integration constant $g(V_g)$ must be set so that the
zero level of $I(\varphi)$ is correct. The CPRs at gate charge
values of $n_g=0$ and $1 e$ are displayed in Fig. \ref{fig:CPR}.
We observe that far away from the charge degeneracy point, $n_g
\sim 0$, the CPR is nearly sinusoidal, while close to $n_g = 1$
the CPR becomes strongly non-sinusoidal. The measured behavior at
$n_g=0$ agrees quite well with theoretical curves calculated using
$E_J/k_B=1.5$ K. The measured variation of $I(\varphi)$ with gate
charge is found to be 10\% stronger than what the theory predicts.
We believe this is partially due to the uncertainty in the parameter values
obtained for $E_J$ and $E_C$ using energy level spectroscopy and,
moreover, to the inaccuracy in the base level of $1/L_J$, which
leads to an accumulation of error in the integration process.

Excitation of the SCPT from the ground state $E_0$ to the first excited level $E_1$ is another possible cause for the discrepancy between measured and calculated CPRs. We note that if the contribution of the upper level is included in simple terms (\emph{i.e.} linear population-weighted addition of $1/L_J$-values of the ground and excited states), then the correction would even worsen the agreement between theory and experiment. This is because the contribution of the level $E_1$ is smaller than that of the ground state, leading to reduced $1/L$ and its integral. Furthermore, inclusion of even higher bands yields similar conclusions as their values for $1/L_J$ are also small.

Effective energy bands with stronger curvature can be obtained if one assumes that the total energy is of the form $\mathcal{H}_S(\varphi) = p_0(\varphi) E_0(\varphi) + p_1(\varphi) E_1(\varphi)$ and that the populations $p_0(\varphi)$ and $p_1(\varphi)$ depend on $\varphi$ on the time scale of $1/f_m$. Then the derivatives of $p_0(\varphi)$ and $p_1(\varphi)$ may make substantial modifications in the inductance and, thereby,  in the integrated CPR. For a simple description, we assume a thermal distribution at effective system temperature $T$ that governs the population of the upper state in the driven system according to
\begin{equation}
  p_1 (n_g,\varphi)= 1-p_0 (n_g,\varphi) = \left(1+e^{\frac{\Delta}{k_bT}} \right)^{-1}
\end{equation}
where $\Delta (n_g,\varphi)= E_1 (n_g,\varphi) - E_0 (n_g,\varphi)$. Below we will restrict ourselves only to a qualitative analysis by assuming that $T$ can be regarded as a constant fitting parameter over the region of interest, i.e. for the data measured at $n_g = 1$.

Using this model, we get for the modified CPR $I_S(\varphi)=\frac{2\pi}{\Phi_0}\frac{d \mathcal{H}_S}{d\varphi}$(bias parameter $n_g=1$e dropped)
\begin{equation}\label{TOTcurr}
I_S (\varphi)=  I^0 (\varphi) + \frac{\partial \Delta ( \varphi)}{\partial \varphi}\left((1-p_0) -\frac{\Delta}{k_bT}p_0^2e^{-\frac{\Delta}{k_bT}}\right)
\end{equation}
where $I^0 (\varphi)$ is the CPR for the ground state.
We have fitted Eq. (\ref{TOTcurr}) to our data in Fig. \ref{fig:CPR} at $n_g=1$ e and find an improved agreement when taking $T = 0.28$ K; at $\varphi=\pi$ this corresponds to $p_1=0.30$ \cite{NOTE}. This value for $T$ is quite high, well above the cross-over temperature of 1e - 2e periodicity in the gate modulation of our SCPT. Consequently, the bath temperature ought to be modified strongly by LZ transitions caused by the measurement drive. Far away from charge degeneracy, LZ-transitions are not effective and no modification of the CPR is needed in our comparison.

Non-sinusoidal CPRs have been reported in a number of Josephson
junctions \cite{GolubovRMP}. However, in
superconducting Cooper pair transistors, where Coulomb energy plays
an important role, the results on (derivatives of) CPRs have remained few and unpersuasive \cite{IlichevPRB04,Zorin2006Nb,konemannPRB07}. Closest to our work is the $L_J$
measurement of Ref. \onlinecite{konemannPRB07} but their result was distinctly affected by a mixing of the ground and excited states due to nonequilibrium quasiparticles. Quasiparticles were also a problem in the determination of the ground state properties of Nb SCPTs in Ref.
\onlinecite{Zorin2006Nb}, where 1e periodicity in $V_g$
was observed contrary to our clean 2e-periodic modulation. Moreover, Born
and coworkers have employed Josephson inductance to characterize
qubits \cite{IlichevPRB04}, but their ratio of $E_J/E_C~\sim 30
... 2$ was too high to make observations similar to ours.

Finally, we believe that our method for constructing the CPR will
be useful also in studies of other exotic Josephson junctions
where the critical current is small, like S-carbon nanotube-S
junctions \cite{tsuneta}. Typically, critical currents in small
junctions have been measured by applying current bias and
recording switching current distributions \cite{JoyezPRL94}. Even
though it is not easy to accomplish, it has been recently shown
that switching currents can be employed to measure current-phase
relations in break junction samples under circumstances where a
large Josephson junction is patterned in parallel with the break
junction \cite{DellaRoccaPRL07}. Because there is no need for an extra junction, we think that our scheme is better
suited for investigations of novel, fragile junctions where extra
lithographic steps may be problematic.

In conclusion, using the phase and magnitude of reflected
microwave signals, we have experimentally measured the Josephson
inductance in a mesoscopic, superconducting double-tunnel-junction
device, i.e., we have determined the quantity that is dual to the
Josephson capacitance \cite{MikaQcap,Duty05}. We have shown by
integrating this Josephson inductance that the current-phase
relation of a SCPT is strongly non-sinusoidal near its charge
degeneracy point and that substantial tuning by the gate voltage
towards nearly sinusoidal CPR at $n_g \sim 0$ can be reached.

We thank Erkki Thuneberg, Alexander Zorin, Jani Tuorila, Tero Heikkil\"a, Matti Laakso and Mikko Paalanen for fruitful discussions and useful comments. This work was supported by the Academy of Finland and by the EU contract FP6-IST-021285-2.

\section{Appendix}

Scans containing the raw data employed in the analysis:

\begin{figure}[h]
\begin{center}
 \includegraphics[width=18pc]{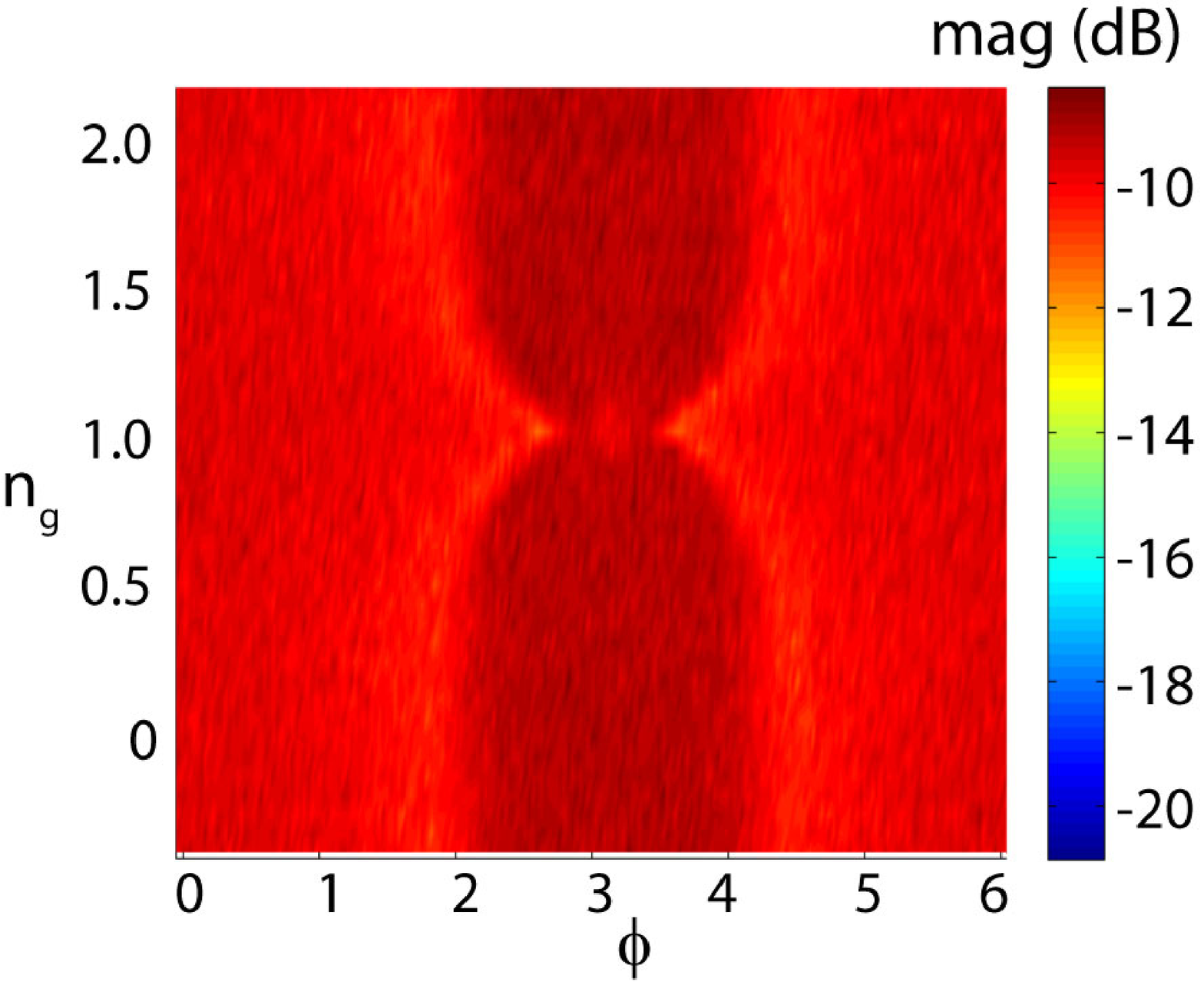}
 \caption{Magnitude of the reflected signal as a function of the phase and charge bias on the superconducting Cooper pair transistor. The color scale in dB is given on the right; the full reflection corresponds to 9.8 dB.}
 \label{fig:mag}
\vspace{2cm}%
 \includegraphics[width=18pc]{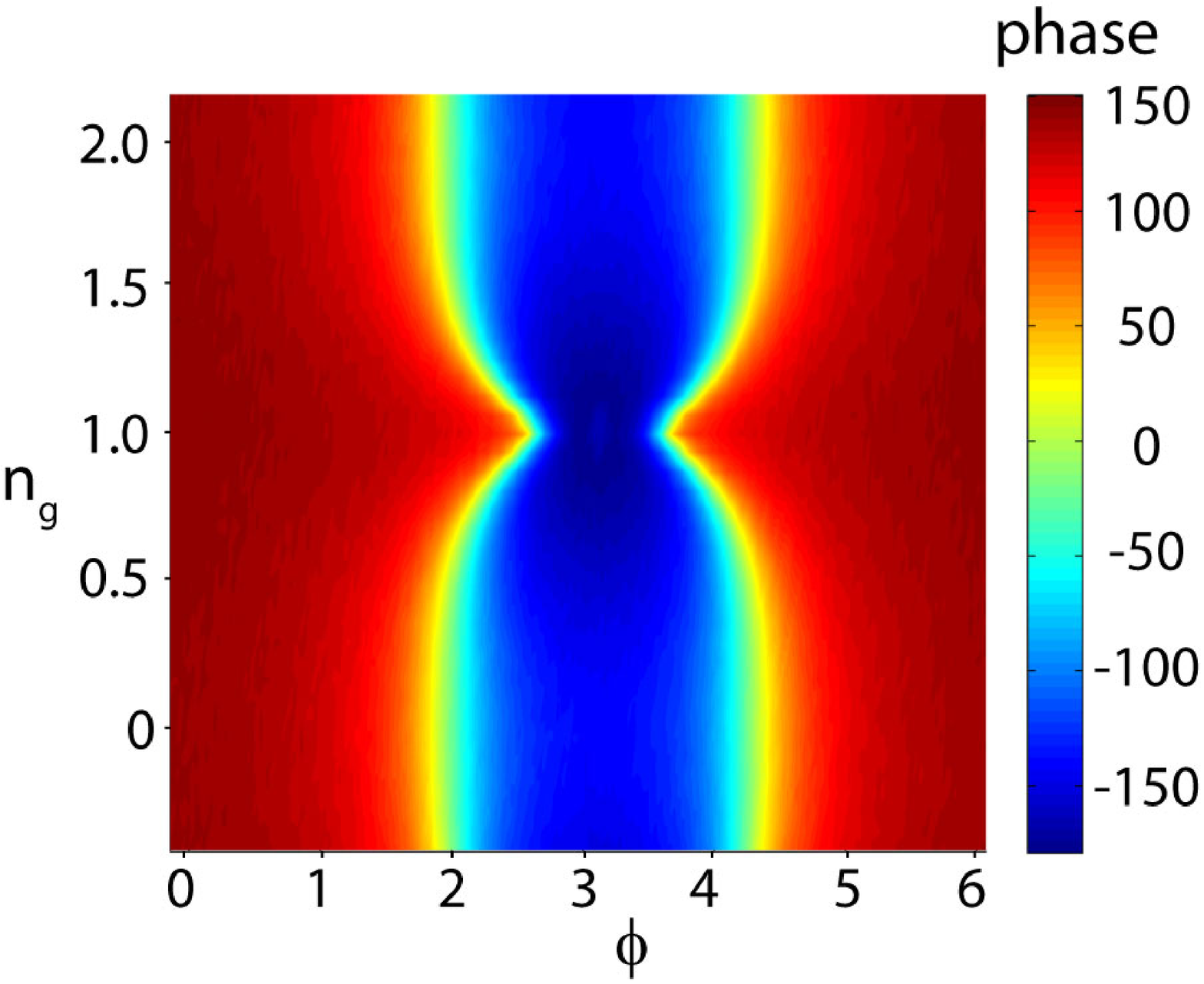}
 \caption{Phase of the reflected signal in degrees as a function of the phase and charge bias. A correction of 25 degrees was made to the reference point of the reflection phase in order to symmetrize the $\Gamma (f)$ on the Smith chart.}
 \label{fig:phase}
\end{center}
\end{figure}


\begin{thebibliography}{30}
\bibitem{NOTE} The worst case estimate for the Landau-Zener probability is $p_{LZ} \sim 0.05$ at the bias point ($n_g$,$\varphi$)=(1,$\pi$); this is sufficient to produce a population of $p_1=0.30$ on the excited band when phase coherent addition of the tunneling amplitudes is accounted for. However, no Landau-Zener intereference phenomena are observed in our data.
\end{thebibliography}
\end{document}